\documentclass[12pt]{article}

\makeatletter

\def\paragraph{\@startsection{paragraph}{4}{\z@}{+2.00ex plus
 +1ex minus +.2ex}{1.5ex plus .2ex}{\it\normalsize}}

\def\section{\@startsection {section}{1}{\z@}{+3.0ex plus +1ex minus
  +.2ex}{2.3ex plus .2ex}{\normalsize\bf\boldmath}}
\def\subsection{\@startsection{subsection}{2}{\z@}{+2.5ex plus +1ex
minus +.2ex}{1.5ex plus .2ex}{\normalsize\bf\boldmath}}
\def\subsubsection{\@startsection{subsubsection}{3}{\z@}{+3.25ex plus
 +1ex minus +.2ex}{1.5ex plus .2ex}{\normalsize\it}}

\expandafter\ifx\csname mathrm\endcsname\relax\def\mathrm#1{{\rm #1}}\fi

\newcounter{saveeqn}

\@addtoreset{equation}{section}

\newcount\@tempcntc
\def\@citex[#1]#2{\if@filesw\immediate\write\@auxout{\string\citation{#2}}\fi
  \@tempcnta\z@\@tempcntb\m@ne\def\@citea{}\@cite{\@for\@citeb:=#2\do
    {\@ifundefined
       {b@\@citeb}{\@citeo\@tempcntb\m@ne\@citea
        \def\@citea{,\penalty\@m\ }{\bf ?}\@warning
       {Citation `\@citeb' on page \thepage \space undefined}}%
    {\setbox\z@\hbox{\global\@tempcntc0\csname
b@\@citeb\endcsname\relax}%
     \ifnum\@tempcntc=\z@ \@citeo\@tempcntb\m@ne
       \@citea\def\@citea{,\penalty\@m}
       \hbox{\csname b@\@citeb\endcsname}%
     \else
      \advance\@tempcntb\@ne
      \ifnum\@tempcntb=\@tempcntc
      \else\advance\@tempcntb\m@ne\@citeo
      \@tempcnta\@tempcntc\@tempcntb\@tempcntc\fi\fi}}\@citeo}{#1}}

\def\@citeo{\ifnum\@tempcnta>\@tempcntb\else\@citea
  \def\@citea{,\penalty\@m}%
  \ifnum\@tempcnta=\@tempcntb\the\@tempcnta\else
   {\advance\@tempcnta\@ne\ifnum\@tempcnta=\@tempcntb \else
\def\@citea{--}\fi
    \advance\@tempcnta\m@ne\the\@tempcnta\@citea\the\@tempcntb}\fi\fi}

\def\nln{\nn\\*[-1ex]}

\def\asymp#1%
{\mathrel{\raisebox{-.4em}{$\widetilde{\scriptstyle #1}$}}}

\def\Nequal#1%
{\mathrel{\raisebox{-.5em}{$\stackrel{=}{\scriptstyle\rm#1}$}}}
\newcommand{\dsl}[1]{\not \hspace{-0.7mm}#1}
\def\dsl{\mathpalette\make@slash}
\def\make@slash#1#2{\setbox\z@\hbox{$#1#2$}%
  \hbox to 0pt{\hss$#1/$\hss\kern-\wd0}\box0}

\def\beq{\begin{equation}}
\def\eeq{\end{equation}}
\def\beqar{\begin{eqnarray}}
\def\eeqar{\end{eqnarray}}
\def\barr#1{\begin{array}{#1}}
\def\earr{\end{array}}
\def\bfi{\begin{figure}}
\def\efi{\end{figure}}
\def\btab{\begin{table}}
\def\etab{\end{table}}
\def\bce{\begin{center}}
\def\ece{\end{center}}
\def\nn{\nonumber}

\def\text{\textstyle}

\def\de{\delta}

\def\si{\sigma}

\def\refeq#1{\mbox{(\ref{#1})}}
\def\refeqs#1{\mbox{(\ref{#1})}}
\def\refeqf#1{\mbox{(\ref{#1})}}

\def\refse#1{\mbox{Section~\ref{#1}}}

\def\citere#1{\mbox{Ref.~\cite{#1}}}
\def\citeres#1{\mbox{Refs.~\cite{#1}}}

\newcommand{\re}{{\mathrm{e}}}
\newcommand{\ri}{{\mathrm{i}}}
\newcommand{\rd}{{\mathrm{d}}}

\newcommand{\rT}{{\mathrm{T}}}

\newcommand{\M}{{\cal M}}
\newcommand{\N}{{\cal N}}
\newcommand{\D}{{\cal D}}
\newcommand{\F}{{\cal F}}
\newcommand{\W}{{\cal W}}
\newcommand{\ST}{{\cal S}}

\def\mathswitchr#1{\relax\ifmmode{\mathrm{#1}}\else$\mathrm{#1}$\fi}

\newcommand{\PW}{\mathswitchr W}
\newcommand{\PZ}{\mathswitchr Z}
\newcommand{\PA}{\mathswitchr A}

\newcommand{\Pd}{\mathswitchr d}

\newcommand{\Pu}{\mathswitchr u}
\newcommand{\Pubar}{\bar{\mathswitchr u}}

\newcommand{\Pt}{\mathswitchr t}

\def\mathswitch#1{\relax\ifmmode#1\else$#1$\fi}

\newcommand{\MW}{\mathswitch {M_\PW}}

\newcommand{\MZ}{\mathswitch {M_\PZ}}

\newcommand{\Pw}{\mathrm w}
\newcommand{\sw}{s_\Pw}
\newcommand{\cw}{c_\Pw}

\newcommand{\ie}{i.e.}

\newcommand{\cf}{cf.\ }

\newcommand{\ct}{{\mathrm{ct}}}

\newcommand{\rL}{{\mathrm{L}}}
\newcommand{\rR}{{\mathrm{R}}}
\newcommand{\qL}{{q,\rL}}
\newcommand{\qR}{{q,\rR}}
\newcommand{\qLR}{{q,\rL/\rR}}
\newcommand{\uL}{{\Pu,\rL}}
\newcommand{\dL}{{\Pd,\rL}}

\newcommand{\dx}{\rd^4 x\;}
\newcommand{\Gcl}{\Gamma_{\mathrm{cl}}}
\newcommand{\Yw}{Y_\Pw}
\newcommand{\Iw}{I_\Pw}
\newcommand{\MSbar}{\overline{\mathrm{MS}}}
\newcommand{\Zbar}{\overline{Z}}

\newcommand{\X}{X}
\newcommand{\U}{\mathrm{U}}
\newcommand{\SU}{\mathrm{SU}}
\newcommand{\BRS}{\mathrm{BRS\mbox{-}inv}}
\newcommand{\inv}{\mathrm{inv}}
\newcommand{\finite}{{\mathrm{fin}}}
\renewcommand{\loop}{{\mathrm{loop}}}
\newcommand{\diag}{{\mathrm{diag}}}
\newcommand{\rig}{{\mathrm{rig}}}
\newcommand{\Vah}{\tilde{V}}
\newcommand{\Uah}{\tilde{Y}}
\newcommand{\Um}{{Y}}

\def\Retilde{\mathop{\widetilde{\mathrm{Re}}}\nolimits}
\def\Imtilde{\mathop{\widetilde{\mathrm{Im}}}\nolimits}

\def\brs{\mathbf s}

\def\pslash#1{{\setbox0=\hbox{$#1$}
  \rlap{\ifdim\wd0>.7em\kern.22\wd0\else\kern.1\wd0\fi /}#1}}

\def\esl{\pslash \varepsilon}
\def\dsl{\pslash \partial}

\def\draftdate{\relax}
\def\mda{\relax}
\def\mua{\relax}
\def\mla{\relax}
\def\draft{
\def\thtystars{******************************}
\def\sixtystars{\thtystars\thtystars}
\typeout{}
\typeout{\sixtystars**}
\typeout{* Draft mode!
         For final version remove \protect\draft\space in source file *}
\typeout{\sixtystars**}
\typeout{}
\def\draftdate{\today}
\def\mua{\marginpar[\boldmath\hfil$\uparrow$]%
                   {\boldmath$\uparrow$\hfil}%
                    \typeout{marginpar: $\uparrow$}\ignorespaces}
\def\mda{\marginpar[\boldmath\hfil$\downarrow$]%
                   {\boldmath$\downarrow$\hfil}%
                    \typeout{marginpar: $\downarrow$}\ignorespaces}
\def\mla{\marginpar[\boldmath\hfil$\rightarrow$]%
                   {\boldmath$\leftarrow $\hfil}%
                    \typeout{marginpar: $\leftrightarrow$}\ignorespaces}
\def\Mua{\marginpar[\boldmath\hfil$\Uparrow$]%
                   {\boldmath$\Uparrow$\hfil}%
                    \typeout{marginpar: $\uparrow$}\ignorespaces}
\def\Mda{\marginpar[\boldmath\hfil$\Downarrow$]%
                   {\boldmath$\Downarrow$\hfil}%
                    \typeout{marginpar: $\downarrow$}\ignorespaces}
\def\Mla{\marginpar[\boldmath\hfil$\Rightarrow$]%
                   {\boldmath$\Leftarrow $\hfil}%
                    \typeout{marginpar: $\leftrightarrow$}\ignorespaces}
\overfullrule 5pt
\oddsidemargin -15mm
\marginparwidth 29mm
}

\def\stars{\strut\leaders\hbox{*}\hfill\strut}
\def\starline{\hfil\strut\hfil\hbox to \textwidth {\stars}\hfil}

\begin{document}
\thispagestyle{empty}
\def\thefootnote{\fnsymbol{footnote}}
\setcounter{footnote}{1} \null \draftdate \strut\hfill
BN-TH-01-2004
\\ \strut\hfill MPP-2004-21\\ \strut\hfill PSI-PR-04-01\\
\strut\hfill hep-ph/0402130 \vfill
\begin{center}
{\Large \bf\boldmath
Physical renormalization condition for the quark-mixing matrix
\par}
\vspace{1cm}
{\large
{\sc A.\ Denner$^1$, E.\ Kraus$^2$ and M.\ Roth$^3$} } \\[1cm]
$^1$ {\it Paul Scherrer Institut, W\"urenlingen und Villigen\\
CH-5232 Villigen PSI, Switzerland} \\[0.5cm]
$^2$ {\it Physikalisches Institut, Universit\"at Bonn, Nu\ss{}allee 12\\
D-53115 Bonn, Germany}
\\[0.5cm]
$^3$ {\it Max-Planck-Institut f\"ur Physik
(Werner-Heisenberg-Institut) \\
D-80805 M\"unchen, Germany}
\par \vskip 1em
\end{center}\par
\vskip 2cm {\bf Abstract:} \par
We investigate the renormalization of the quark-mixing matrix in the
Electroweak Standard Model. 
The corresponding counterterms are gauge independent as can be 
shown using an extended BRS symmetry.  
Using rigid $\SU(2)_\rL$ symmetry, we prove that the
ultraviolet-divergent parts of the invariant counterterms are related
to the field renormalization constants of the quark fields.  We point
out that for a general class of renormalization schemes rigid
$\SU(2)_\rL$ symmetry cannot be preserved in its classical form, but
is renormalized by finite counterterms.  Finally, we discuss a genuine
physical renormalization condition for the quark-mixing matrix that is
gauge independent and does not destroy the symmetry between quark
generations.
\par
\vskip 2cm \noindent February 2004
\null \setcounter{page}{0}
\clearpage
\def\thefootnote{\arabic{footnote}}
\setcounter{footnote}{0}

\section{Introduction}
\label{se:intro}

Presently, the parameters of the quark-mixing matrix (QMM) are being
precisely measured at the B factories.  When calculating precision
observables involving the QMM, in general the renormalization of the
QMM is required. This was first realized in \citere{Marciano:cn} for
the Cabibbo angle in the Standard Model (SM) with two fermion
generations.  An example where the counterterms for the QMM for three
generations have been taken into account can be found in
\citere{Gambino:1998rt}. In the SM the effects of the renormalization
of the QMM are numerically small, since the masses of all down-type
quarks are small compared to the W-boson mass \cite{Denner:1990yz}.
However, a consistent renormalization of the QMM should be formulated
for conceptual reasons. Moreover, the renormalization of mixing
matrices may become phenomenologically relevant in extensions of the
SM.

The most straight-forward way to renormalize the QMM is to directly
fix the four independent parameters of the QMM, three angles and a
CP-violating phase, by choosing four suitable observables, e.g.\ four
specific W-boson decays \cite{Barroso:2000is}. However, the
counterterms determined in this way depend on the chosen observables,
and the symmetry between the amplitudes involving different
generations is destroyed. A symmetric renormalization condition can be
obtained naturally using the modified minimal subtraction ($\MSbar$)
scheme (see e.g.\ \citeres{Balzereit:1998id,Pilaftsis:2002nc}).  This,
however, is not a physical condition and depends on an arbitrary
renormalization scale.  Moreover in this scheme, the renormalized
$S$-matrix elements exhibit singularities of the form
$1/(m_{q,i}^2-m_{q,k}^2)$ in the limit of degenerate up-type or
down-type quark masses $m_{q,i}\approx m_{q,k}$, i.e., the limit of
degenerate quark masses, where the QMM is equal to the unit matrix and
need not be renormalized, is not approached smoothly.

A renormalization condition for the QMM in the on-shell scheme was
first proposed in \citeres{Denner:1990yz,Denner:1993kt}. In this
proposal, the  counterterms of the QMM are determined from the
field renormalization constants of the quark fields in the
on-shell renormalization scheme. This prescription is simple, does
not introduce a renormalization scale, and is smoothly connected
to the limit of degenerate quark masses. Later it was discovered
\cite{Gambino:1998ec}, however, that the renormalization condition
of \citeres{Denner:1990yz,Denner:1993kt} leads to
gauge-parameter-dependent counterterms for the QMM and thus to
gauge-parameter-dependent $S$-matrix elements. In
\citere{Gambino:1998ec} a modified renormalization condition was
proposed based on field renormalization constants defined at zero
momentum. This scheme gives gauge-parameter-independent results at
the one-loop level, but leads to singularities in the $S$-matrix
elements for degenerate quark masses.  Moreover, it is not clear
whether it can be generalized beyond one-loop order.

It was also suggested to split off the gauge-parameter-dependent
part of the on-shell quark-field renormalization constants as far
as the definition of the QMM counterterm is concerned, i.e., to
define the QMM counterterm from the quark-field renormalization
constants calculated in the `t~Hooft--Feynman gauge
\cite{Yamada:2001px}. This scheme corresponds exactly to the
original one of  \citeres{Denner:1990yz,Denner:1993kt}. It is
gauge-parameter independent by definition, but of course depends
implicitly on the choice of the `t~Hooft--Feynman gauge.
Generalizing this philosophy, it was argued in
\citere{Pilaftsis:2002nc} that any renormalization scheme for the
QMM may be viewed as a gauge-invariant scheme by definition. This
is possible since any scheme is related to the (gauge-invariant)
$\MSbar$ scheme by ultraviolet(UV)-finite matrices which can be
chosen to match any renormalization condition.

In \citere{Diener:2001qt} desirable properties for the renormalization
condition of the QMM have been formulated. These are UV
finiteness, gauge-parameter independence, and unitarity of the
renormalized QMM. In
addition, the renormalization condition should be physically motivated
and treat all generations on an equal footing. The requirement that
the renormalized amplitudes approach the limit of degenerate up-type
or down-type masses smoothly is also implicitly contained in this
paper. A
renormalization condition was formulated
that obeys all these properties and is also applicable to the lepton
mixing in Majorana-neutrino theories. In this scheme, the renormalized
QMM is fixed by matching the matrix elements for W-boson decay in the
SM with those in reference theories with zero mixing and different
assignments of down-type quarks to the generations. The unitarity of
the renormalized QMM is obtained by subtracting the
unitarity-violating part from the counterterm obtained from the
reference theory.

All the mentioned prescriptions have only been used at the
one-loop level, and it is not clear how they can be consistently
generalized to higher orders. Recently, a renormalization
prescription for the QMM has been proposed \cite{Zhou:2003gb} that
could overcome all these weaknesses. This renormalization
condition has been introduced via a two-step procedure, and
\citere{Zhou:2003gb} leaves a lot of questions open. In the
present paper we rederive the renormalization condition of
\citere{Zhou:2003gb} in a different way and put it on a more sound
basis.

Before we consider explicit renormalization conditions for the QMM
we first investigate the consequences of the symmetries of the
theory on the QMM and its renormalization.  In gauge theories, the
gauge-parameter dependence of Green functions can be controlled by
extending the gauge parameter $\xi$ to a Becchi--Rouet--Stora (BRS)
doublet ($\xi,\chi$),
where $\chi$ is a Grassmann-valued parameter. Gauge-parameter
dependence of Green functions and counterterms is determined by an
extended Slavnov--Taylor (ST) identity \cite{Kluberg-Stern:rs,Piguet:1984js}.
By solving the extended ST identity it is seen that, in general, physical
parameters and their counterterms have to be gauge-parameter
independent.  Finally, it is possible to prove gauge-parameter
independence of physical $S$-matrix elements
\cite{Piguet:1984js,Kummer:2001ip}.

This formalism has been first applied to the renormalization of
the QMM in \citere{Gambino:1998ec} yielding the result that
counterterms to the QMM are gauge-parameter independent. As an
additional constraint the authors of \citere{Gambino:1998ec}
require that the Ward--Takahashi identity of gauge invariance in
the background-field gauge be preserved in its classical form to
all orders.  As we show, the gauge-parameter dependence of
$S$-matrix elements and counterterms is governed by the BRS
invariance only. Thus, the use of the Ward--Takahashi identity is
not adequate in this context. 
In particular, invariance of the Ward--Takahashi identity
implies that the renormalization of the QMM is related to the
renormalization of quark fields. If the renormalization of the QMM
is required to be gauge independent in this context, the
renormalization of the quark fields must be gauge independent,
too.  Thus, a complete on-shell renormalization scheme is not
admissible in this approach. Therefore, the authors of
\citere{Gambino:1998ec} modify the on-shell conditions for quark
fields to renormalizations at zero momentum, at least as far as
the renormalization of the QMM is concerned.

In the present paper we investigate the consequences of the
symmetries of the theory on the QMM  and its renormalization in
the general linear $R_\xi$ gauge.  As already mentioned, the
relevant symmetry to control the gauge-parameter dependence of
counterterms as well as of physical $S$-matrix elements is BRS
symmetry in form of the ST identity.  On the other hand, rigid
invariance restricts the UV divergences of the invariant
counterterms and, in particular, relates the divergent parts of
the field renormalization constants and of the counterterms of the
QMM.  However, finite field redefinitions can be introduced that
renormalize the Ward identity of rigid invariance. These are, in
particular, needed in the complete on-shell scheme in order to
have enough freedom to introduce complete on-shell conditions in
agreement with gauge-parameter independence of the QMM.

This paper is organized as follows: in \refse{se:BRS} we show that the
counterterms for the QMM are gauge-parameter independent as a consequence
of the BRS invariance of the SM. Using rigid symmetry, we prove
in \refse{se:rigid} that the UV-divergent parts of the invariant
counterterms of the QMM are related to the field renormalization
constants of the quark fields.
The free parameters of the QMM and their counterterms are elaborated
in \refse{se:QMMparameters}. Finally, we discuss a physical
renormalization condition for the QMM in \refse{se:rencond}. The
appendix contains some discussion of absorptive parts.

\section{Implications of BRS invariance on counterterms}
\label{se:BRS}

The interplay between the renormalization of the QMM and the field
renormalization of the quark fields makes it particularly difficult to
disentangle the gauge-parameter dependence of the different kinds of
counterterms. The relevant symmetry that governs the gauge-parameter
dependence is the BRS symmetry. This is considered in this section.

Following closely the conventions of \citere{Denner:1993kt} the part
of the classical action of the SM relevant for quark mixing reads
\begin{eqnarray}
\label{eq:Gcl}
\Gcl^{\mathrm{quark}}
&= &\int \dx \Bigg\{\ri \bar Q_i^\rL \gamma_\mu D_{ij}^{\mu} Q_j^\rL
+ \ri \bar u_i^\rR \gamma_\mu D_{ij}^{\mu} u_j^\rR
+ \ri \bar d_i^{\,\rR} \gamma_\mu D_{ij}^{\mu} d_j^\rR
\nonumber \\ && {}
- m_{\Pd,i} (\bar d_i^{\,\rL} d_i^\rR +\bar d_i^{\,\rR} d_i^\rL)
- m_{\Pu,i} (\bar u_i^{\rL} u_i^\rR +\bar u_i^\rR u_i^\rL)
\nonumber \\ && {}
- \frac {e} {\sqrt{2} \MW \sw}\left[
  \bar Q_i^\rL {\mathbf V}_{ij} \Phi m_{\Pd,j}  d_j^\rR
+ \bar Q_i^\rL {\mathbf V}^\dagger_{ij} (\ri \sigma^2) \Phi^{*} m_{\Pu,j}
u_j^\rR + \mbox{h.c.}\right]\Bigg\},
\end{eqnarray}
where the left-handed quarks (isospin doublets) are denoted by
$Q^\rL_i=(\Pu^\rL_i,\Pd^\rL_i)^\rT$, the right-handed quarks
(isospin singlets) by $\Pu^\rR_i,\Pd^\rR_i$, and the Highs doublet,
with vacuum expectation value $(0,v/\sqrt{2})^\rT$ subtracted, by
$\Phi$.
The indices $i,j$ run over $N$ quark families.  The sine and cosine of
the weak mixing angle are defined as usual by
$\cw=\sqrt{1-\sw^2}=\MW/\MZ$, and $\si^2$ is a Pauli matrix.

The matrix ${\mathbf V}$ is a unitary $2N \times 2N $ matrix in
the quark-family and isospin space and is composed as
\begin{equation}
{\mathbf V}_{ij} =
\left(\begin{array}{cc}
V_{ij} & 0 \\
0 & \delta _{ij}
\end{array}\right), \qquad
{\mathbf V}^\dagger {\mathbf V} = {\mathbf 1}.
\end{equation}
This matrix includes the QMM $V_{ij}$ which is a unitary $N \times N$
matrix depending on $N(N -1)/2$ angles and $(N-1)(N-2)/2$ phases.
These are the Cabibbo angle for $N =2$ and the three
angles and one phase of the Cabibbo--Kobayashi--Maskawa matrix for $N=3$.
In case of CP conservation all phases vanish.

The covariant derivatives of the quark fields read
\begin{eqnarray}
\label{eq:covder}
D^\mu_{ij} Q_j^\rL &=&
  \bigg\{ \delta_{ij} \partial^\mu
+ \ri \frac{e}{\cw}  \delta_{ij}\frac{\Yw^{Q}}{2}
  \left(\sw Z^\mu + \cw A^\mu \right)
\nonumber \\ && {}
- \ri \frac{e}{\sw}\left[V_{ij} \Iw^+ W^{+\mu}
+ V_{ij}^\dagger \Iw^- W^{-\mu}
+ \delta_{ij} \Iw^3 (\cw Z^\mu - \sw A^\mu)\right]\bigg\}Q^\rL_j,
\nonumber \\
D^\mu_{ij} q_j^\rR &=&
  \bigg[\partial^\mu
+ \ri \frac{e}{\cw} \frac{\Yw^{q}}{2}
  \left(\sw Z^\mu + \cw A^\mu \right) \bigg] \delta_{ij} q^\rR_j,
\qquad q^\rR_j=u^\rR_j,d^\rR_j,
\end{eqnarray}
where $V_{ij}^\dagger=V_{ji}^*$, and the generators of the
$\SU(2)_\rL$ gauge group are defined by $\Iw^a=\sigma^a/2$ with the
Pauli matrices $\sigma^a$.
For convenience, we replaced in \refeq{eq:covder} the generators
$\Iw^{1,2}$ by
\begin{equation}
\Iw^\pm=\frac{1}{\sqrt{2}}\left(\Iw^1\pm \ri \Iw^2\right).
\end{equation}
The hypercharges of the fields $Q^\rL_j$ and $q^\rR_j$ are denoted by
$\Yw^{Q}$ and $\Yw^{q}$, respectively.  Note that the QMM $V_{ij}$
appears in \refeq{eq:covder} only in terms involving W bosons.

The classical action, partially given by \refeq{eq:Gcl}, is
complete in the sense that it is the most general field polynomial
that is consistent with power counting and respects all symmetry
requirements of the underlying theory. In  particular, invariance
under BRS symmetry implies unitarity of the QMM.  Besides BRS
symmetry, rigid $\SU(2)_\rL$ symmetry and local $\U(1)_Y$ gauge
symmetry of hypercharge are the relevant symmetries of the SM (see
\citere{Kraus:1997bi}). Since the BRS symmetry controls the
gauge-parameter dependence of the QMM, we focus on BRS
transformations in the following and come
back to gauge transformations later.

The BRS transformations of the quark fields take the form
\begin{eqnarray}
\brs Q_i^\rL &=&
\bigg\{
- \ri \frac{e}{\cw} \delta_{ij} \frac{\Yw^Q}{2}
  \left( \sw c_\PZ + \cw c_\PA \right)
\nonumber \\ && {}
+ \ri \frac{e}{\sw} \left[V_{ij} \Iw^+ c_+
+ V_{ij}^\dagger \Iw^- c_-
+ \delta_{ij} \Iw^3 \left(\cw c_\PZ -\sw c_\PA \right)
\right] \bigg\} Q_j^\rL,
\nonumber \\
\brs q_i^\rR &=&
- \ri \frac{e}{\cw} \frac{\Yw^q}{2}
  \left(\sw c_\PZ + \cw c_\PA\right) \delta_{ij} q_j^\rR,
\end{eqnarray}
where $c_\PA$, $c_\PZ$, and  $c_\pm$ are Faddeev--Popov ghost fields.
The BRS transformations for vector-boson and scalar fields have the usual
form and can be found, for instance, in \citeres{Kraus:1997bi,Kraus:1998wb}.

Since the BRS transformations of the quark fields are composite operators
and receive quantum corrections, we couple them
to external fields $\psi^\rL_{q,i}$ and $\Psi^\rR_{i}$,
\begin{equation}
\label{eq:Gext}
\Gcl^{\mathrm{ext}} = \int \dx \left[
  \ldots
+ \sum_{i=1}^{N} \left(
  \bar\psi_{\Pu,i}^\rL \brs u_i^\rR
+ \bar\psi_{\Pd,i}^\rL \brs d_i^\rR
+ \bar\Psi_i^\rR \brs Q_i^\rL
+ \mbox{h.c.} \right)\right],
\end{equation}
and add $\Gcl^{\mathrm{ext}}$ to the classical action. The auxiliary
non-propagating fields $\psi^\rL_{q,i}$ and
$\Psi^\rR_{i}=(\psi^\rR_{\Pu,i},\psi^\rR_{\Pd,i})^\rT$ have ghost
charge $-1$ and are BRS invariants, i.e., $\brs
\psi^{\rL/\rR}_{q,i}=0$.  For quantization, BRS transformations are
encoded in the Slavnov--Taylor (ST) identity
\begin{eqnarray}
\label{eq:STfermion}
\ST(\Gamma) &=&
\int \dx \Bigg[\ldots + \sum_{i=1}^{N} \Bigg(
  \frac{\delta \Gamma}{\delta \bar\psi^\rL_{\Pu,i}}
  \frac{\delta \Gamma}{\delta u^\rR_i}
+ \frac{\delta \Gamma}{\delta \bar\psi^\rL_{\Pd,i}}
  \frac{\delta \Gamma}{\delta d^\rR_i}
+ \frac{\delta \Gamma}{\delta \bar\Psi^\rR_i}
  \frac{\delta \Gamma}{\delta Q^\rL_i}
+ \hbox{h.c.} \Bigg) \Bigg]=0,
\end{eqnarray}
where $\Gamma$ is the generating functional of one-particle
irreducible Green functions. The ellipses 
in \refeq{eq:Gext} and
\refeq{eq:STfermion} denote the contributions from vector-boson,
scalar, lepton, and ghost fields (see
\citeres{Kraus:1997bi,Kraus:1998wb} for details). As usual, the
linearized ST operator $\brs_\Gamma$ is defined by the expansion
\begin{equation}
\ST(\Gamma+\Delta) = \ST(\Gamma) + \brs_\Gamma \Delta + {\cal O}(\Delta^2).
\end{equation}

All counterterms compatible with the ST identity are called {\it
  BRS-invariant} counterterms in the following.  These comprise
counterterms that are invariant under rigid symmetry, but also those
that are not (see \refse{se:rigid}).
According to the definition of the BRS-invariant counterterms, they are
invariant under the linearized ST operator,
\begin{equation}
\brs_{\Gcl} \Gamma_{\BRS,i} = 0.
\end{equation}
These counterterms can be expressed in form of differential operators
$\D_i$ that commute with the linearized ST operator,
\begin{equation}
\label{eq:CTBRS}
\Gamma_{\BRS,i} = \delta Z_i(\xi) \D_i \Gcl,\qquad
\D_i \ST(\F)-\brs_{\F} \D_i {\F} = 0,
\end{equation}
where ${\cal F}$ is an arbitrary field polynomial.  In order to
investigate the gauge-parameter dependence of the renormalization
constants $\delta Z_i(\xi)$, we introduce BRS-varying gauge
parameters following \citeres{Kluberg-Stern:rs,Piguet:1984js},
\begin{equation}
\brs \xi = \chi,  \qquad  \brs \chi = 0,
\end{equation}
where $\xi$ is a gauge parameter, and the new auxiliary constant field
$\chi$ is Grassmann valued and has ghost charge $-1$.
We include the BRS transformation of $\xi$ into an extended
ST identity,
\begin{equation}
\label{eq:SText}
\ST^\chi(\Gamma ) = \ST(\Gamma)
+ \chi \partial_\xi \Gamma=0.
\end{equation}
This construction allows us to classify the counterterms into genuine
gauge-parameter-de\-pen\-dent and -independent ones, and to prove
the gauge-parameter independence of the $S$-matrix elements (see
\citere{Kummer:2001ip} for details).  It is important to realize
that the introduction of the additional constant field $\chi$ is
only an auxiliary construction and does not affect the
$\chi$-independent part of $\Gamma$, since the part of $\Gamma$
involving the BRS doublet $(\xi,\chi)$  is free of anomalies by
construction, and $\chi$ does not appear in ST identities for
physical Green functions.

In the following, we are searching for counterterms that are invariant
with respect to the extended ST identity \refeq{eq:SText},
\begin{equation}
\brs_{\Gcl}^\chi \Gamma_{\BRS,i}^{\chi}=0,
\end{equation}
and investigate the consequences for the gauge-parameter dependence of
the renormalization constants $\delta Z_i(\xi)$. In general, the
counterterms \refeq{eq:CTBRS} violate the extended ST
identity,
\begin{equation}
\brs_{\Gcl}^\chi \Gamma_{\BRS,i}=
\chi \left(\partial_\xi \ln \delta Z_i(\xi)\right)
\Gamma_{\BRS,i}.
\end{equation}
There are two possibilities to construct counterterms compatible with
the extended ST identity:
\begin{enumerate}
\item
If a local field polynomial $\hat\Delta_i$ exists such that
\begin{equation}
\label{eq:BRSvar}
\Gamma_{\BRS,i}=\delta Z_i(\xi) \brs_{\Gcl} \hat\Delta_i,
\end{equation}
we are able to build an invariant counterterm for the extended
ST identity \refeq{eq:SText} by defining
\begin{equation}
\Gamma_{\BRS,i}^{\chi}:=\Gamma_{\BRS,i}
- \chi \left(\partial_\xi \delta Z_i(\xi)\right) \hat\Delta_i.
\end{equation}
In this case there is no restriction on the gauge-parameter dependence of
the renormalization constant $\delta Z_i(\xi)$. Typical examples of
such counterterms are field renormalizations of the matter fields,
gauge-fixing and ghost terms, and in the case of the Minimal Supersymmetric
Standard Model soft-supersymmetry-breaking terms if they are
introduced according to \citere{Hollik:2002mv}.
\item If $\Gamma_{\BRS,i}$ cannot be written in form of
  \refeq{eq:BRSvar}, the renormalization constant $\delta Z_i$ must be
  gauge-parameter independent, and the counterterm of the extended ST
  identity reads
\begin{equation}
\label{eq:extended}
\Gamma_{\BRS,i}^{\chi}:=\Gamma_{\BRS,i}
\qquad \mbox{with} \qquad
\partial_\xi \delta Z_i=0.
\end{equation}
Examples of such counterterms are those for the physical
parameters of the SM, like the gauge couplings and masses
(cf.~\citeres{Haussling:1996rq,Haussling:1997ix,Haussling:1998pp,Gambino:1999ai}).
\end{enumerate}
As a result, the counterterms split into two classes: those of
the first class can be written as a $\brs_{\Gcl}$-variation and are in
general gauge-parameter dependent;
those of the second class cannot be written in the
form of \refeq{eq:BRSvar} and thus must be genuinely gauge-parameter
independent,
provided that appropriate renormalization
conditions are chosen that do not lead to an artificial gauge-parameter
dependence.  A careful separation between both classes is necessary in
order to draw conclusions on gauge-parameter dependence.

For the renormalization of the QMM the following counterterms are
relevant:
\begin{enumerate}
\item
The field renormalizations of the quark fields result from
$\brs_{\Gcl}$-variations as
\begin{equation}
\label{eq:fieldrenorm}
\Gamma_{\BRS,ij}^{q^{\rL/\rR}}=-\delta Z_{ij}^\qLR
\brs_{\Gcl} \left(\bar\psi_{q,i}^{\rR/\rL} q_j^{\rL/\rR}\right)
+ \mbox{h.c.}
= \delta Z_{ij}^\qLR \N^\qLR_{ij} \Gcl + \mbox{h.c.}
\end{equation}
with
\begin{equation}
\N^\qLR_{ij}=\int \dx \left[
\Big(q_j^{\rL/\rR}\Big)^\rT \frac{\delta}{\delta q_i^{\rL/\rR}}
-\bar\psi_{q,i}^{\rR/\rL}
\frac{\delta}{\delta\Big(\bar\psi_{q,j}^{\rR/\rL}\Big)^\rT}\right].
\end{equation}
Thus, the field renormalization constants of the quark fields are
in general gauge-parameter dependent, which is a well-known fact.
The field renormalization constants $Z^\qLR$ are in general
complex $N \times N$ matrices.
The field-number operator $\N^\qLR_{ij}$ commutes with the
ST operator [c.f.\ \refeq{eq:CTBRS}].

Equivalently, field renormalization can be introduced
using the field redefinitions
\begin{eqnarray}
\label{eq:fieldred}
q^{\rL/\rR}_i &\to& Z^\qLR_{ij} q^{\rL/\rR}_j
= \left(\delta_{ij} + \delta Z^\qLR_{ij}\right) q^{\rL/\rR}_j,
\nonumber \\
\bar\psi_{q,i}^{\rR/\rL}&\to&
\bar\psi_{q,j}^{\rR/\rL} (Z^\qLR)^{-1}_{ji}
=\bar\psi_{q,j}^{\rR/\rL}
\left(\delta_{ji} - \delta Z^\qLR_{ji}\right),
\end{eqnarray}
and the one obtained by taking the hermitian adjoint.
The ST operator is invariant under these transformations.
However, the rigid transformations are not invariant under
\refeq{eq:fieldred} as will be stressed in the next section.
\item
On the other hand the counterterms corresponding to the parameters
 $\theta_n$ of the QMM and the quark masses $m_{q,k}$ cannot
be written in form of a $\brs_{\Gcl}$-variation. Therefore, they are
genuinely gauge-parameter-independent quantities. These counterterms read
\begin{equation}
\label{eq:optype2}
\Gamma_{\BRS}^{\theta_n}=
\delta \theta_n \frac{\partial}{\partial \theta_n} \Gcl,\qquad
\Gamma_{\BRS}^{m_{q,k}}=\delta m_{q,k} \frac{\partial}{\partial m_{q,k}} \Gcl,
\end{equation}
and can be introduced by parameter redefinitions,
\begin{equation}
\theta_n\to \theta_n+\delta \theta_n, \qquad
m_{q,k}\to m_{q,k}+\delta m_{q,k}
\end{equation}
with
\begin{equation}
\partial_\xi \delta \theta_n=
\partial_\xi \delta m_{q,k}=0.
\end{equation}
\end{enumerate}

\section{Restrictions from rigid $\SU(2)_\rL$ invariance}
\label{se:rigid}

In this section, we want to construct all invariant counterterms of the SM 
relevant for our case. Invariant counterterms correspond to local field 
operators that respect the defining symmetries of the underlying model. 
The relevant symmetries for the renormalization of the QMM in the SM are 
the extended BRS symmetry as discussed in the previous section but also the 
(spontaneously broken) rigid $\SU(2)_\rL$ gauge symmetry. The latter is 
usually expressed in form of a Ward identity. In the following, we construct 
the invariant counterterms from the BRS-invariant counterterms by requiring 
rigid $\SU(2)_\rL$ gauge symmetry in addition. Like in the case of the 
BRS-invariant counterterms [\cf \refeq{eq:CTBRS}], we define
the invariant counterterms by invariant operators that commute with both 
the ST and Ward operators. Finally, we introduce finite field 
redefinitions for the quark fields, which leads to a renormalization of 
$\SU(2)_\rL$ gauge symmetry. We show that these new parameters are required 
to impose on-shell renormalization conditions for the quark fields.

Before discussing the invariant counterterms, we would like to 
comment on symmetry breaking resulting from the chosen 
regularization. If we assume an invariant regularization scheme for 
the renormalization in the SM we need only invariant counterterms 
for the renormalization procedure. Unfortunately for the SM 
there is no regularization method known which respects all symmetries
owing to the so-called $\gamma_5$ problem. Therefore, we require that 
the symmetry breaking owing to the regularization be restored in a first 
step by introducing symmetry-restoring counterterms. These counterterms 
need not respect the symmetries of the underlying model and can also 
include UV divergences. 
In the following, we assume that this has been done and the symmetries are 
restored. Hence, the remaining UV divergences respect the symmetry 
identities and can be absorbed by invariant counterterms only. 

The rigid $\SU(2)_\rL$ Ward identities take the form
\begin{equation}
\label{eq:rigidWard}
\W_a \Gamma=
\int \dx \delta^{\rig}_a \phi_k \, \frac{\delta \Gamma}{\delta \phi_k}
=0+{\cal O}(\chi), \qquad a=1,2,3,4,
\end{equation}
where $\phi_k$ runs over all fields.  The Ward operators $\W_a$,
$a=1,2,3$, in (\ref{eq:rigidWard}) respect the algebra of the
$\SU(2)_\rL$ group,
\begin{equation}
[\W_a,\W_b]=\ri\epsilon_{abc} \W_c,\qquad a,b,c=1,2,3,
\end{equation}
and commute with the Ward operator $\W_4$ of $\U(1)_Y$ symmetry of
hypercharge.  The Ward identity has to be satisfied by the
$\chi$-independent part of $\Gamma$, while the unphysical part
involving the field $\chi$ need not be rigidly invariant.  Using
the definition
\begin{equation}
\W_\pm=\frac{1}{\sqrt{2}}\left(\W_1\pm \ri \W_2\right),
\end{equation}
in the classical approximation the rigid transformations for
$\W_\pm$ take  the form
\begin{equation}
\label{eq:rigid}
\begin{array}[b]{r@{\,}l@{\quad}r@{\,}l@{\quad}r@{\,}l@{\quad}r@{\,}l}
\delta_+^\rig u^\rL_i &= \frac{\ri}{\sqrt{2}}
V_{ij} d_j^\rL, &
\delta_+^\rig \bar u_i^\rL &= 0, &
\delta_-^\rig u_i^\rL &= 0, &
\delta_-^\rig \bar u_i^\rL &= - \frac{\ri}{\sqrt{2}}
\bar d_j^{\,\rL} V_{ji}^\dagger, \\
\delta_+^\rig d_i^\rL &= 0, &
\delta_+^\rig \bar d_i^{\,\rL} &= - \frac{\ri}{\sqrt{2}}
\bar u_j^\rL V_{ji}, &
\delta_-^\rig d^\rL_i &= \frac{\ri}{\sqrt{2}}
V_{ij}^\dagger u_j^\rL, &
\delta_-^\rig \bar d_i^\rL &= 0, \\
\delta_+^\rig \psi_{\Pu,i}^\rR &= \frac{\ri}{\sqrt{2}}
V_{ij} \psi_{\Pd,j}^\rR, &
\delta_+^\rig \bar\psi_{\Pu,i}^\rR &= 0, &
\delta_-^\rig \psi_{\Pu,i}^\rR &= 0, &
\delta_-^\rig \bar\psi_{\Pu,i}^\rR &= - \frac{\ri}{\sqrt{2}}
\bar\psi_{\Pd,j}^\rR
V_{ji}^\dagger,\\
\delta_+^\rig \psi_{\Pd,i}^\rR &= 0, &
\delta_+^\rig \bar\psi_{\Pd,i}^\rR &= - \frac{\ri}{\sqrt{2}}
\bar\psi_{\Pu,j}^\rR V_{ji},
&
\delta_-^\rig \psi_{\Pd,i}^\rR &= \frac{\ri}{\sqrt{2}}
V_{ij}^\dagger \psi_{\Pu,j}^\rR, &
\delta_-^\rig \bar\psi_{\Pd,i}^\rR &= 0.
\end{array}
\end{equation}
The rigid transformations for the auxiliary fields $\psi^\rR_q$
are defined such that the ST operator is invariant under
\refeq{eq:rigid}. The rigid transformations of the other fields
take their usual form (see e.g.\ \citere{Kraus:1998wb}).

Counterterms that respect both the ST identity {\em and\/} the Ward
identities \refeq{eq:rigidWard} of rigid $\SU(2)_\rL$ symmetry in the
classical form (\ref{eq:rigid}) are called {\it invariant}
counterterms.  Like in \refse{se:BRS}, we discuss the
gauge-parameter-dependent and -independent counterterms separately.
\begin{enumerate}
\item In order to generate quark-field counterterms invariant under
  BRS {\em and\/}
  rigid symmetries, we search for all field
  redefinitions \refeq{eq:fieldred} that leave the rigid Ward
  identities
  invariant.  While the Ward operators
  $\W_{3,4}$ are not affected by the replacements \refeq{eq:fieldred},
  $\W_\pm$ in (\ref{eq:rigid})  are in general modified by \refeq{eq:fieldred}.  Requiring that the rigid transformations \refeq{eq:rigid} be
  invariant under
  the replacement \refeq{eq:fieldred}, the field renormalizations
  \refeq{eq:fieldred} of the left-handed fields are restricted by
  $(Z^\uL_{\inv})^{-1} V Z^\dL_{\inv}=V$, resulting in
\begin{equation}
\label{eq:restriction}
Z^\dL_{\inv}=V^\dagger Z^\uL_{\inv} V.
\end{equation}
Since the field renormalizations of the right-handed fields
are not restricted by rigid invariance, the corresponding
BRS-invariant
counterterms are also invariant counterterms, i.e., $Z^\qR_{\inv}=Z^\qR$.
Furthermore, the operators $\N^\qLR_{ij}$ corresponding to the invariant
counterterms $Z^\qLR_{\inv}$ commute with the Ward operators.

\item
The renormalization operators $\delta m_{q,k} \partial/\partial m_{q,k}$
of \refeq{eq:optype2} commute with the ST and Ward operators and,
hence, generate invariant counterterms,
\begin{equation}
\Gamma_{\inv}^{m_{q,k}}=\delta m_{q,k} \frac{\partial}{\partial m_{q,k}} \Gcl.
\end{equation}

On the other hand, the operators that correspond to the
renormalization of the angles and phases of the QMM do not commute
with the Ward operators $\W_\pm$. In order to disentangle the
counterterms to the QMM from the counterterms of the field
renormalizations, we construct the invariant operators that
commute with both the ST operator and the Ward operators as
\begin{eqnarray}
\label{eq:thetactop}
\D_{\theta_n} &=&
  \frac{\partial}{\partial\theta_n}
+   \Bigg[\frac{1}{2} \N^\uL_{ij} \frac{\partial V_{ik}}{\partial \theta_n}
    V^\dagger_{kj}
  - \frac{1}{2}\N^\dL_{ij} V^\dagger_{ik}
    \frac{\partial V_{kj}}{\partial \theta_n}
  +\mbox{h.c.}
\Bigg].
\end{eqnarray}
These operators define the gauge-parameter-independent, invariant
counterterms to the QMM,
\begin{equation}
\Gamma_{\inv}^{\theta_n}= \delta\theta_n \D_{\theta_n} \Gcl,
\end{equation}
where the parameters $\theta_n$ run over the angles and phases of the
QMM.  This renormalization includes both renormalization
transformations of the mixing angles and field renormalizations,
\begin{eqnarray}
\label{eq:thetanren}
\theta_n&\to& \theta_n+\delta \theta_n,
\nonumber \\
u^\rL_i &\to& \left[\delta_{ij}
    + \frac{1}{2} \delta \theta_n
\left(\partial_{\theta_n} V_{ik}\right) V^\dagger_{kj}\right] u^\rL_j,
\nonumber \\
d^\rL_i &\to& \left[\delta_{ij}-\frac{1}{2}\delta \theta_n
    V^\dagger_{ik} \left(\partial_{\theta_n} V_{kj}\right) \right]
    d^\rL_j,
\nonumber \\
\bar\psi_{\Pu,i}^\rR&\to& \bar\psi_{\Pu,j}^\rR
    \left[\delta_{ji}- \frac{1}{2}\delta \theta_n
    \left(\partial_{\theta_n} V_{jk}\right) V^\dagger_{ki} \right],
\nonumber \\
\bar\psi_{\Pd,i}^\rR&\to& \bar\psi_{\Pd,j}^\rR
    \left[\delta_{ji}+ \frac{1}{2}\delta \theta_n
    V^\dagger_{jk} \left(\partial_{\theta_n} V_{ki}\right) \right],
\end{eqnarray}
and the corresponding hermitian adjoint transformations.

Both $\delta m_{q,k}$ and $\delta \theta_n$ are genuine
gauge-parameter-independent counterterms.
\end{enumerate}
The relation \refeq{eq:restriction} and the renormalization of the
QMM \refeq{eq:thetanren} restrict the UV-divergence structure of the
invariant counterterms.

The remaining field renormalization parameters of \refeq{eq:fieldred},
\ie, those that do not respect \refeq{eq:restriction}, belong to a new
type of renormalization constants, namely finite field redefinitions.
\begin{enumerate}
\setcounter{enumi}{2}
\item
We introduce finite field redefinitions for the left-handed
down-type  quarks since $\delta Z^\dL_{\inv}$ is constrained by
\refeq{eq:restriction}.  The finite field redefinitions of the
down-type quarks compatible with the ST identity read
\begin{eqnarray}
\label{eq:finitefieldred}
d^\rL_i &\to& R^{\finite}_{ij} d^\rL_j
= \left(\delta_{ij} + \delta R^{\finite}_{ij}\right) d^\rL_j,
\nonumber \\
\bar\psi_{\Pd,i}^\rR&\to&
\bar\psi_{\Pd,j}^{\rR} (R^{\finite})^{-1}_{ji}
= \bar\psi_{\Pd,j}^\rR
\left(\delta_{ji} - \delta R^{\finite}_{ji}\right)
\end{eqnarray}
with an arbitrary complex $N\times N$ matrix $\delta R^{\finite}$.
Since the replacement \refeq{eq:finitefieldred} is done everywhere,
i.e., in the action, in the ST operator, and in the Ward
operators, it does not disturb the validity of the symmetry requirements
of the theory.
While the ST operator and $\W_{3,4}$ stay unchanged,
the renormalized rigid transformations corresponding to $\W_\pm$
are modified. The renormalized Ward operators $\W_\pm$ are obtained
from \refeqs{eq:rigid} by the substitution \refeq{eq:finitefieldred}.
As shown in \refse{se:BRS}, these field redefinitions are in general
gauge-parameter dependent. Furthermore,
these renormalization constants do not include UV divergences and are
only needed to satisfy on-shell renormalization conditions.
\end{enumerate}

With the so-defined invariant counterterms and finite field redefinitions,
we are able to define a more convenient set of renormalization constants,
\begin{eqnarray}
\label{eq:fieldred2}
V_{ij}&\to&V_{ij} + \delta V_{ij},
\nonumber\\
q^{\rL/\rR}_i&\to&Z_{ij}^\qLR q^{\rL/\rR}_j
=\left(\delta_{ij}+\delta Z_{ij}^\qLR\right)q^{\rL/\rR}_j,
\nonumber\\
\bar\psi^{\rR/\rL}_i&\to&\bar\psi^{\rR/\rL}_j
(Z^\qLR)_{ji}^{-1}
=\bar\psi^{\rR/\rL}_j\left(\delta_{ji}-\delta Z_{ji}^\qLR\right)
\end{eqnarray}
with
\begin{eqnarray}
\label{eq:physCT}
\delta V&=&\delta \theta_n \left(\partial_{\theta_n} V\right),
\nonumber\\
\delta Z^\uL(\xi)&=&\delta Z^\uL_{\inv}(\xi)
+  \frac{1}{2}\delta \theta_n
  \left(\partial_{\theta_n} V\right) V^\dagger,
\nonumber\\
\delta Z^\dL(\xi)&=&V^\dagger \delta Z^\uL_{\inv}(\xi) V
-  \frac{1}{2}\delta \theta_n V^\dagger
  \left(\partial_{\theta_n} V\right)
+ \delta R^{\finite}(\xi),
\nonumber\\
\delta Z^\qR(\xi)&=&\delta Z^\qR_{\inv}(\xi).
\end{eqnarray}
 In \refeq{eq:physCT} we have indicated the gauge-parameter
dependence of the renormalization constants explicitly. The
definition of the renormalization constant $\delta V$ in
\refeq{eq:physCT} implies that the renormalized QMM stays unitary
in all orders by construction as it is required by BRS invariance
of the theory.

Using \refeq{eq:physCT} we can express the UV
divergences of the QMM in terms of UV divergences of left-handed
field redefinitions:
\begin{eqnarray}
\label{eq:divergent} \delta V &=&\delta Z^\uL V - V \delta Z^\dL
+\mbox{finite terms}, \nonumber\\ \delta V&=&-(\delta
Z^\uL)^\dagger V + V (\delta Z^\dL)^\dagger +\mbox{finite terms},
\end{eqnarray}
where we used $(\partial_{\theta_n} V) V^\dagger=-V
(\partial_{\theta_n} V^\dagger)$. As the UV divergences satisfy
also the extended ST  identity, the gauge-parameter dependent part
of field redefinitions cancels in \refeq{eq:divergent} in such a
way that the UV divergences of the QMM are gauge independent.
Since in the $\MSbar$ scheme only the UV-divergent parts and
gauge-independent constants are subtracted, $\delta V$ can be
consistently determined by $\MSbar$ field renormalizations,
\begin{equation}
\label{eq:MS}
\delta V^{\MSbar}\stackrel{!}{=}-\frac{1}{2}\left\{
\left[(\delta Z^{\uL,\MSbar})^\dagger
-\delta Z^{\uL,\MSbar}\right] V
+ V \left[\delta Z^{\dL,\MSbar}
-(\delta Z^{\dL,\MSbar})^\dagger\right]\right\}.
\end{equation}
From \refeq{eq:physCT} it can be seen that the renormalization scheme
which uses an $\MSbar$ subtraction for $\delta V$ but on-shell
conditions for $\delta Z^\qLR$ is fully consistent
\cite{Balzereit:1998id,Pilaftsis:2002nc} yielding, however, contrary
to the pure $\MSbar$ scheme renormalized Ward identities of rigid
symmetry.

In \citeres{Denner:1990yz,Denner:1993kt},  the
relation \refeq{eq:MS} is used as a
renormalization condition also for the finite parts,%
\footnote{Note that our notation differs from the one of
\citere{Denner:1993kt} by a factor of $2$ in the definition of the
field renormalization constants. }
\begin{equation}
\label{eq:DennerSack}
\delta V\stackrel{!}{=}-\frac{1}{2}\left\{
\left[(\delta Z^\uL)^\dagger
-\delta Z^\uL\right] V
+ V \left[\delta Z^\dL
-(\delta Z^\dL)^\dagger\right]\right\},
\end{equation}
which is equivalent to
\begin{equation}
\delta R^\finite - (\delta R^\finite)^\dagger \stackrel{!}{=}0.
\end{equation}
 This renormalization condition sets the anti-hermitian
part of the renormalization constants $\delta R^\finite(\xi)$ to zero.
In order to satisfy the on-shell conditions, $\delta Z^\uL$ and
$\delta Z^\dL$ are needed as independent counterterms at our disposal
(see \refse{se:QMMparameters}).
As \refeq{eq:physCT} shows, the gauge-dependent parts of $\delta
Z^\uL(\xi)$ and $\delta Z^\dL(\xi)$ can in general be absorbed in
$\delta Z^\uL_{\inv}(\xi)$ and $ \delta R^{\finite}(\xi)$.
However, if $\delta R^\finite - (\delta R^\finite)^\dagger=0$, the
renormalization condition \refeq{eq:DennerSack} requires one 
to adjust $\delta \theta_n$. This leads in general to a
gauge-parameter-dependent $\delta \theta_n$, which is inconsistent
with the extended ST identity.  For this reason the
renormalization condition \refeq{eq:DennerSack} yields
gauge-parameter-dependent results for physical matrix elements, as
has been confirmed by an explicit one-loop calculation in
\citere{Gambino:1998ec}.

To circumvent these problems, the authors of
\citere{Gambino:1998ec} give up the complete on-shell conditions,
but propose renormalization conditions respecting the Ward
identity \refeq{eq:rigidWard} in its classical form, i.e., $\delta
R^{\finite}=0$ and resulting in gauge-parameter-independent
counterterms to QMM
by using fermion-field
renormalization constants fixed at zero momentum.
In this scheme, gauge independence of the QMM has been confirmed by
an explicit one-loop calculation, but could not be confirmed  to
all orders.

In \citere{Gambino:1998ec} the motivation for constructing a scheme 
in accordance with the classical Ward identity is based on the
theorem that any renormalization prescription that preserves the
rigid Ward identity in its classical form
leads to a gauge-parameter-independent definition of the QMM. 
Note that this theorem requires the assumptions that 
the extended ST identity is satisfied and that the Ward operator commutes 
with the extended ST operator, as can be seen in the proof of this 
theorem in \citere{Gambino:1998ec}. However, in this form the theorem 
of \citere{Gambino:1998ec} is only of little use, especially in the 
on-shell scheme where the Ward identity has to be renormalized as 
discussed before.
Preserving the rigid Ward identity means to take  $\delta
R^\finite(\xi)=0$. Then, the rigid Ward identity is preserved in
its classical form for any $\delta \theta_n$ irrespective of its
gauge-parameter dependence. 
As we have shown in \refse{se:BRS}, it is the extended  BRS invariance 
alone which controls gauge-parameter dependence or independence of Green 
functions and $S$-matrix elements, and which yields in the present case the
gauge independence of counterterms to the QMM.

For this reason we do not establish conditions  which are
motivated by rigid invariance like \refeq{eq:DennerSack} 
or the scheme of  \citere{Gambino:1998ec} but fix the counterterms of
the QMM directly on physical matrix elements (see \refse{se:rencond}). 

\section{Parameters of the quark-mixing matrix}
\label{se:QMMparameters}

Before we turn to the definition of a physical renormalization condition for
the QMM in \refse{se:rencond}, we investigate the different types
of parameters included in the QMM and the field redefinitions $Z^\qLR$,
and study how these parameters contribute to the bilinear action and to the
 $\PW^+\bar\Pu_i {\Pd}_j$ vertex.

The SM allows the generalized field redefinitions \refeq{eq:fieldred2} of the
quark fields where $Z^\qLR$ are general complex $N\times N$ matrices.
A general complex matrix can be decomposed into a hermitian and
a unitary matrix,
\begin{equation}
\label{eq:decompglobal}
Z^\qLR = U^\qLR H^\qLR
\end{equation}
with
\begin{equation}
(H^\qLR)^\dagger = H^\qLR,
\qquad (U^\qLR)^\dagger U^\qLR = {\mathbf 1}.
\end{equation}

Applying the field redefinitions \refeq{eq:fieldred2} to the bilinear
part of the classical action \refeq{eq:Gcl} yields
\begin{equation}
\Gamma_{\mathrm{bil}}^{\mathrm{field\mbox{-}red}}=
\int \dx \sum_{q=u,d} \left[
  \ri \bar{q}_i^{\rL} \X^\qL_{ij} \dsl q_i^{\rL}
+ \ri \bar q_i^{\rR} \X^\qR_{ij} \dsl q_i^{\rR}
+ \left(\bar{q}_i^{\rL} M^q_{ij} q_j^{\rR} + \mbox{h.c.}\right)
\right]
\end{equation}
with
\begin{equation}
\X^\qLR = (H^\qLR)^\dagger H^\qLR, \qquad
M^q = (H^\qL)^\dagger (U^\qL)^\dagger M^q_\diag U^\qR H^\qR,
\end{equation}
and $M^q_\diag=\diag(m_{q,1},\ldots,m_{q,N})$.

Inspecting this result, we see that the hermitian parts of the
matrices $Z^\qL$ and $Z^\qR$ can be determined on kinetic terms, while
the unitary parts can be fixed on mass terms up to a common complex
diagonal matrix $\diag(\exp(\ri\tilde
  \varphi^{q}_1),\ldots,\exp(\ri\tilde \varphi^{q}_N))$, which can be
 extracted as
\begin{eqnarray}
\label{decomposition} U_{ij}^{q,\rL} = \sum_{k=1}^{N}
\re^{\ri\tilde \varphi^{q}_i} \delta _{ik}
 \tilde U^{q,\rL}_{kj},
\qquad U_{ij}^{q,\rR} = \sum_{k=1}^{N} \re^{\ri\tilde
\varphi^{q}_i} \delta _{ik}
 \tilde U^{q,\rR}_{kj}.
\end{eqnarray}
The diagonal matrix can be parametrized as
\begin{equation}\label{eq:unphysphases}
\re^{\ri\tilde \varphi^{q}_i} \delta_{ij} =
\left(\re^{\ri\sum_{n=1}^{N-1} \varphi^{q}_n T^\diag_n}
  \re^{\ri\varphi^{q}_0 T_0^\diag}\right)_{ij},
\end{equation}
where the matrices $T^\diag_n$ denote the $N-1$ traceless
diagonal generators of $\SU(N)$
and $T_0^\diag={\mathbf 1}/\sqrt{2N}$.

The number of physical parameters of the QMM for $N$ quark
families is determined as follows: BRS invariance  implies that
the QMM is a general unitary matrix. A general unitary matrix
$\Um$ has $N^2$ parameters. We can use the diagonal matrices that
are not fixed on bilinear terms to remove $2N-1$ phases from $\Um$
and obtain the usual form of the QMM including only physical
parameters,
\begin{eqnarray}
\label{eq:UinV}
V &=& \re^{- \ri\sum_{n=1}^{N-1}\varphi^{\Pu}_n T^\diag_n}
\re^{- \ri\varphi^{\Pu}_0 T_0^\diag} \Um
\re^{\ri\sum_{n=1}^{N-1}\varphi^{\Pd}_n T^\diag_n}
\re^{\ri\varphi^{\Pd}_0 T_0^\diag}
\nonumber \\
&=& \re^{- \ri( \varphi^{\Pu}_0 - \varphi^{\Pd}_0) T_0^\diag}
\re^{- \ri\sum_{n=1}^{N-1}\varphi^{\Pu}_n T^\diag_n} \Um
\re^{\ri\sum_{n=1}^{N-1}\varphi^{\Pd}_n T^\diag_n},
\end{eqnarray}
where we used the fact  that $T_0^\diag$ is the only generator
that commutes with all generators of the $\SU(N)$ group. Thus, the
QMM has $N^2-(2N-1)= (N-1)^2$ physical parameters.  These consist
of $N(N-1)/2$ angles $\vartheta_m$ of an $N\times N$ orthogonal
matrix and of $(N-1) (N-2)/2$ complex phase factors
$\exp(\ri\varphi_l)$.

Similar to the decomposition \refeq{eq:decompglobal} of the field
redefinition matrices, the corresponding counterterms $\delta
Z^\qLR$ can be written as a linear combination of hermitian and
anti-hermitian matrices,
\begin{equation}
\delta Z^\qLR = \sum _{n=0}^{N^2-1} \de z_n^\qLR T_n,
\end{equation}
where $\de z_n^\qLR$ are complex numbers and $T_n$, $n=1,\ldots,N^2-1$ are the
generators of $\SU(N)$ and  $T_0={\mathbf 1}/\sqrt{2N}$.
In the same way as we have discussed for the matrices $Z^\qLR$ in the
beginning of this section, not all of the $\de z_n^\qL$ and $\de
z_n^\qR$
can be determined on the bilinear terms,
\begin{equation}
\delta\Gamma_{\mathrm{bil}}^{\mathrm{field\mbox{-}red}}=
\int \dx \sum_{q=u,d} \left[
  \ri \bar{q}_i^{\rL} \delta \X^\qL_{ij} \dsl q_i^{\rL}
+ \ri \bar q_i^{\rR} \delta \X^\qR_{ij} \dsl q_i^{\rR}
+ \left(\bar{q}_i^{\rL} \delta M^q_{ij} q_j^{\rR} + \mbox{h.c.}\right)
\right]
\end{equation}
with
\begin{equation}
\delta \X^\qLR = (\delta Z^\qLR)^\dagger + \delta Z^\qLR, \qquad
\delta M^q = (\delta Z^\qL)^\dagger M^q_\diag
+ M^q_\diag \delta Z^\qR.
\end{equation}
Common imaginary parts of the coefficients $\de z_n^\qL$ and $\de z_n^\qR$
corresponding to the diagonal generators $T_n^\diag$ remain as free
parameters. Splitting off these free parameters, we obtain
\begin{equation}\label{eq:decompZ}
\delta Z^\qL = \delta \tilde{Z}^\qL
+ \ri \sum_{n=0}^{N-1} c^q_n T_n^\diag, \qquad
\delta Z^\qR = \delta \tilde{Z}^\qR
+ \ri \sum_{n=0}^{N-1} c^q_n T_n^\diag
\end{equation}
with free real parameters $c_n^q$, and $\delta \tilde{Z}^\qL$ and
$\delta \tilde{Z}^\qR$ having fixed values for the imaginary parts of
the diagonal entries. One can choose for example real diagonal entries
for the left-handed field renormalization constants
$\delta \tilde{Z}^\qL_{ii}= (\delta \tilde{Z}^{\qL}_{ii})^*$.

Now we turn to the counterterms of the QMM. Owing to BRS
invariance  $(V+\delta V)$ is a unitary matrix. The unitarity
constraint reads
\begin{equation}
\label{eq:unitary}
(V + \delta V) (V + \delta V)^\dagger
= (V + \delta V)^\dagger (V + \delta V) = {\mathbf 1},
\end{equation}
which implies
\begin{equation}
\delta V V^\dagger + V \delta V^\dagger = - \delta V \delta V^\dagger
,\qquad
\delta V^\dagger V + V^\dagger \delta V = - \delta V^\dagger \delta V.
\end{equation}
These equations are implicitly solved by decomposing $\delta V V^\dagger$
into a hermitian and an anti-hermitian part,
\begin{equation}
\label{eq:solutiongeneralV}
\delta V = - \frac{1}{2} \delta V \delta V^\dagger V + \delta \Vah
\qquad \mbox{with} \qquad
\delta \Vah V^\dagger = - \left(\delta \Vah V^\dagger\right)^\dagger.
\end{equation}
Equation \refeq{eq:solutiongeneralV} can be solved perturbatively for
$\delta V$ once $\delta \Vah$ is given. Therefore, $\delta \Vah$ has
the same number of independent parameters as $\delta V$ or $V$,
namely $(N-1)^2$.

In order to formulate a renormalization condition for the QMM,
we investigate the $\PW^+\Pubar_i\Pd_j$ vertex.
Including counterterms, this vertex reads
\begin{equation}
\Gamma_{\PW\bar\Pu\Pd}=
\frac{e}{\sqrt{2}\sw}
\int \dx \bar u_i^\rL \gamma^\mu W^+_\mu (V+\de F_{\ct})_{ij} d_j^\rL
\end{equation}
 with the matrix \beq \label{eq:Wudvert} \de F_{\ct}=
V\left(\delta Z_\PW +\frac{\delta e}{e}
  -\frac{\delta \sw}{\sw}\right)
+ (\delta Z^{ \uL})^\dagger V + V \delta Z^{ \dL} + \delta V.
\eeq
Inserting the decomposition \refeq{eq:decompZ} as well as
\refeq{eq:solutiongeneralV}, this can be written as
\beq
\label{eq:Wudvert2}
\de F_{\ct}= V\left(\delta Z_\PW +\frac{\delta e}{e}
  -\frac{\delta \sw}{\sw} \right)
  + (\delta \tilde{Z}^\uL)^\dagger V + V \delta \tilde{Z}^\dL
- \frac{1}{2} \delta V \delta V^\dagger V + \delta \Uah,
\eeq
where we defined
\begin{equation}
\label{eq:deltaU}
\delta \Uah = - \ri \sum_{n= 1}^{N^2-1}
\left(c^\Pu_n T_n^\diag V - V   c^\Pd_n T_n^\diag\right)
- \ri \left(c^\Pu_0 - c^\Pd_0\right) V
+ \delta \Vah.
\end{equation}
While  $\de Z_\PW$, $\de e$, $\de\sw$, and $\de \tilde Z^{q,\rL}$ are
fixed from other vertex functions, we have the $N^2$ parameters of the
matrix $\delta \Uah$ at our disposal for renormalization conditions of
the QMM: $(N-1)^2$ free parameters from $\delta \Vah$, $2(N-1)$ real
constants $c^q_n$ from the traceless diagonal generators $T^\diag_n$,
and one real constant $c^q_0$ from $T^\diag_0$. These parameters are
just sufficient to fix a general unitary matrix.
For later use we note that \refeq{eq:deltaU} and
\refeq{eq:solutiongeneralV} imply the anti-hermiticity of $\delta \Uah V^\dagger$,
\begin{equation}
\label{eq:propertyYtilde}
\delta \Uah V^\dagger = - \left(\delta \Uah V^\dagger\right)^\dagger.
\end{equation}

\section{Physical renormalization of the quark-mixing matrix}
\label{se:rencond}

In the previous section we found that the field renormalization
constant $\delta Z^\qLR$ can be determined on the quark self-energies
only up to some unphysical phases.
These phases can be used to extend the QMM to a general unitary
matrix. In this section we formulate a physical renormalization
condition for this unitary matrix and thus for the QMM.

As we have already seen in \refse{se:BRS}, the counterterms to the QMM
are genuinely gauge-parameter independent.  In order not to
introduce an artificial gauge-parameter dependence, the
renormalization conditions have to be chosen properly.  If
gauge-parameter-independent matrix elements that involve the QMM are
available, these matrix elements can be used to determine
$\delta V$.  If we ignore for the moment the instability of the $\PW$
bosons and the quarks, such matrix elements are those of the decays
$\PW^+ \to \Pu_i \bar\Pd_j$ or $\bar\Pu_i \to \PW^-
\bar\Pd_j$ if a top quark is involved.  Both types of decays are
related by crossing symmetry.

Before we come to the actual renormalization condition, we want to add
some comments on the difficulties related to unstable particles.
Matrix elements to the decays $\PW^+ \to \Pu_i \bar\Pd_j$ or
$\bar\Pu_i \to \PW^- \bar\Pd_j$ suffer from the fact that the external
particles are unstable.  Contrary to stable particles it is not known
how to construct gauge-parameter-independent matrix elements with
unstable particles at external legs. Nevertheless we use these matrix
elements for a renormalization condition for the QMM. The problem
related to the instability of the external particles manifests itself
in contributions of the order of the decay width of these particles,
which we cannot control.  Several attempts to obtain
gauge-parameter-independent matrix elements involving internal or
external unstable particles have been undertaken in the literature
(see e.g.\
\citeres{Veltman:th,Stuart:1995zr,Grassi:2000dz,Espriu:2002xv}).  For
the case where no gauge-parameter-independent matrix element is
available, a fully consistent prescription has been given in
\citere{Piguet:1984js}, which defines renormalization conditions for
gauge-parameter-independent counterterms on arbitrary
gauge-parameter-dependent Green functions.  In this section we treat
the external particles as stable and ignore the problems related to
their finite decay widths.  This approach implies that absorptive
parts should be disregarded in the following results. Some remarks on
the absorptive parts are made at the end of this section and in the
appendix.

Following closely the notation of \citere{Denner:1993kt}, the lowest-order
matrix element for the decay $\PW^+ \to \Pu_i\bar\Pd_j$ reads
\begin{equation}
\M_{0,ij} = V_{ij} \M_{1,ij}^-
\end{equation}
with the standard matrix element
\begin{equation}
\label{eq:SME}
\M^-_{1,ij}=- \frac{e}{\sqrt{2} \sw} \bar{u}(p_{\Pu,i}) \esl(p_{\PW})
\omega_- v(p_{\Pd,j}),
\end{equation}
and the chiral projector $\omega_-=(1-\gamma_5)/2$.
The matrix element including radiative corrections can be written as
\begin{equation}
\label{eq:Mcorrections}
\M_{ij} =
\sum_{a=1}^2 \sum_{\sigma=\pm} F_{a,ij}^{\sigma} \M_{a,ij}^\sigma
\end{equation}
with four standard matrix elements
$\M_a^\sigma$. The corresponding four gauge-invariant form factors
$F_a^\sigma$ are functions of $\MW^2$, $m_{\Pu,i}^2$, and
$m_{\Pd,j}^2$.  The form factor $F_{1}^-$, the only one appearing at
lowest order and thus involving overall UV divergences and
counterterms, can be decomposed as
\begin{equation}
\label{eq:formfactor}
F_1^- = V +\delta F_{\loop,1}^{-}
+ \delta F_{\ct}= V+\sum_{l\ge 1} \left(\delta F_{\loop,1}^{-\,(l)}
+ \delta F_{\ct}^{(l)}\right),
\end{equation}
where $\delta F_{\loop,1}^{-\,(l)}$ summarizes the $l$-th order loop
contributions including also counterterm insertions of lower orders and
$\delta F_{\mathrm{ct}}^{(l)}$ includes the $l$-th order overall counterterms
as defined in \refeq{eq:Wudvert}.

Since the form factor $F^-_1$ depends on the counterterms of the
QMM, it can be used to define a proper renormalization condition
for the QMM.  Owing to loop corrections this form factor is a
general complex $N\times N$
matrix. We decompose%
\footnote{The decomposition $F^-_1 = \Um H'$ leads to the same
 renormalization condition \refeq{eq:CPresult}.}
 this form factor into a unitary matrix $\Um$ and
a hermitian matrix $H$,
\begin{equation}
F^-_1=H \Um \qquad \mbox{with} \qquad
H^\dagger = H, \qquad \Um^\dagger \Um = {\mathbf 1}.
\end{equation}
As discussed at the end of \refse{se:QMMparameters}, we have enough
parameters at our disposal for the renormalization of the QMM to fix a
general unitary matrix. This allows us to require that the unitary
part of the form factor $F^-_1$
not receive quantum corrections in higher orders, \ie,
\begin{equation}
\label{eq:rencond1}
\Um\stackrel{!}{=}V.
\end{equation}
The renormalization condition \refeq{eq:rencond1} can be written as
\begin{equation}
\label{eq:rencond2}
F^-_1 \stackrel{!}{=} H V.
\end{equation}
Using the hermiticity of $H$, we obtain as the final result for the
renormalization condition,
\begin{equation}
\label{eq:rencond3}
F^-_1 V^\dagger - V (F^-_1)^\dagger \stackrel{!}{=} 0
\end{equation}
or using \refeq{eq:formfactor} 
\begin{equation}
 (V+\delta
F_{\loop,1}^{-}+ \delta F_{\ct}) V^\dagger \stackrel{!}{=} V
(V+\delta F_{\loop,1}^{-} + \delta F_{\ct})^\dagger.
\end{equation}
For the $l$-loop contribution this reads
\begin{equation}
\label{eq:rencondl}
(\de F^{-(l)}_{\loop,1}+\de F^{(l)}_{\ct}) V^\dagger \stackrel{!}{=}
V (\de F^{-(l)}_{\loop,1}+\de F^{(l)}_{\ct})^\dagger.
\end{equation}
Inserting \refeq{eq:Wudvert2} and using the anti-hermiticity of
$\de\Uah V^\dagger$ and the fact that $\de e/e$ and $\de\sw/\sw$
are real, this equation can be solved for $\de\Uah^{(l)}$,
\begin{eqnarray}
\label{eq:CPresult}
\delta \Uah^{(l)}  &=&
- \frac{1}{2} \left[(\delta \tilde{Z}^{\uL(l)})^\dagger
  - \delta \tilde{Z}^{\uL(l)} \right] V
- \frac{1}{2} V \left[\delta \tilde{Z}^{\dL(l)}
  - (\delta \tilde{Z}^{\dL(l)})^\dagger \right]
\nonumber \\ && { }
-\frac{1}{2} (\de Z_\PW^{(l)} - \de Z_\PW^{*(l)})
-\frac{1}{2} \left[ \delta F_{\loop,1}^{-(l)}
- V (\delta F_{\loop,1}^{-(l)})^\dagger V \right].
\end{eqnarray}

We can rewrite \refeq{eq:CPresult} into standard form by absorbing the
undetermined phases into the field renormalization and find, dropping
the loop index $l$,
\begin{eqnarray}
\label{eq:CPresultconventional}
\delta \Vah &=&
- \frac{1}{2} \left[(\delta Z^\uL)^\dagger - \delta Z^\uL \right] V
- \frac{1}{2} V \left[\delta Z^\dL - (\delta  Z^\dL)^\dagger \right]
\nonumber \\ && {}
-\frac{1}{2} (\de Z_\PW - \de Z_\PW^{*})
- \frac{1}{2} \left[\delta F_{\loop,1}^- - V (\delta
  F_{\loop,1}^-)^\dagger V \right].
\end{eqnarray} This together with \refeq{eq:solutiongeneralV} determines the
counterterms to the QMM.  As already mentioned, our conventions
differ by a factor of 2 in the definition of the field
renormalization constants from those of \citere{Denner:1993kt}. We
note that \refeq{eq:CPresultconventional} implicitly requires us
to fix the $c_n^q$ such that $\delta V=\delta\theta_n \partial
V/\partial\theta_n$, \ie, the free parameters in $\de Z^{q,\rL}$
must be fixed such that the renormalized QMM can be expressed in
terms of the corresponding renormalized physical parameters 
(see also \citere{Liao:2003jy}).

Let us now discuss the renormalization condition
\refeq{eq:CPresultconventional} leaving aside absorptive parts: it
is a physical renormalization condition that satisfies all
requirements mentioned in the introduction. The corresponding
renormalized QMM is gauge independent, unitary by construction,
and symmetric with respect to the fermion generations. Moreover,
the renormalized matrix elements approach the limit of degenerate
fermion masses smoothly.  An apparent drawback of the
renormalization condition \refeq{eq:CPresultconventional} is that
it requires the calculation of the vertex form factor $\delta
F_{\loop,1}^-$.  This, however, is anyhow needed for all processes
involving the QMM.  The renormalization condition
\refeq{eq:CPresultconventional} is equivalent to the one given by
Zhou in \citere{Zhou:2003gb}.  However, while we impose a
renormalization condition on a physical matrix element
(\ref{eq:rencond3}) that preserves the unitarity of the QMM,
Zhou requires a condition that violates unitarity. In a second
step he corrects this by extracting the unitarity-preserving part
of a counterterm, a procedure that was also used in the
renormalization scheme proposed in \citere{Diener:2001qt}.  
As a further remark,
we note that exactly the same renormalization condition
\refeq{eq:CPresultconventional} is obtained from the decays $\PW^+
\to \Pu_i\bar\Pd_j$ and $\bar\Pu_i \to \PW^- \bar\Pd_j$ or $\PW^-
\to \bar\Pu_i\Pd_j$ and $\Pu_i \to \PW^+ \Pd_j$.

If we take into account absorptive parts, we encounter a number of
problems and drawbacks: the counterterm
\refeq{eq:CPresultconventional} obtained from the decays $\PW^+ \to
\Pu_i\bar\Pd_j$ and $\bar\Pu_i \to\PW^- \bar\Pd_j$ differs from the
one obtained analogously from the decays $\PW^- \to \bar\Pu_i\Pd_j$
and $\Pu_i \to\PW^+ \Pd_j$ owing to \refeq{a:FGrel2}, and the
counterterms to the QMM become complex also in the case of CP
conservation.  Moreover, when absorptive parts of the loop corrections
and of the Lehmann--Symanzik--Zimmermann (LSZ) 
factors are included in the calculation of $S$-matrix
elements, the limit of degenerate fermion masses is no longer
approached smoothly.  This problem is related to the lack of a
consistent definition of $S$-matrix elements for unstable external
particles and applies to all other existing renormalization
prescriptions for the QMM once absorptive parts are taken into
account. One proposal for the modification of the LSZ factors in the
presence of external unstable particles \cite{Espriu:2002xv} and the
renormalization of the QMM in this approach is discussed in some
detail in the appendix.  However, it also does not solve the mentioned
problems.

Therefore, we advocate to only include dispersive parts in
\refeq{eq:CPresultconventional}.

\section{Conclusions}

We have studied the renormalization of the quark-mixing matrix
(QMM) and the corresponding restrictions from BRS invariance and
rigid $\SU(2)_\rL$ symmetry.

We started from the fact that the gauge-parameter dependence of
counterterms and physical $S$-matrix elements can be controlled
using a modified Slavnov--Taylor identity, where the gauge parameter
$\xi$ is extended to a BRS doublet by introducing an auxiliary
field $\chi$.  
Using this formalism, it can be seen that counterterms that
cannot be written as a BRS variation must be genuinely gauge-parameter
independent, the others may be gauge-parameter dependent.
While the
quark field renormalization constants are generally gauge-parameter
dependent, the counterterms to the QMM do not
depend on the gauge parameters if appropriate physical renormalization
conditions are imposed.

In order to satisfy complete on-shell renormalization conditions
in the quark sector, finite gauge-parameter-dependent field
redefinitions must be introduced.  Imposing complete on-shell
conditions without these finite field redefinitions induces an
artificial gauge-parameter dependence in the counterterms to the
QMM, as found by an explicit one-loop calculation
\cite{Gambino:1998ec} for the renormalization prescription of
\citeres{Denner:1990yz,Denner:1993kt}.  These finite field
redefinitions appear explicitly in the rigid $\SU(2)_\rL$
transformations resulting in a renormalization of the rigid
symmetry. From rigid $\SU(2)_\rL$ invariance we found relations
between the ultraviolet-divergent parts of the invariant
counterterms.

Finally, we proposed a physical renormalization condition for the
QMM based on the decays $\PW^+ \to \Pu_i \bar\Pd_j$ and
$\bar\Pt\to\PW^-\bar\Pd_j$. This condition fixes all counterterms
properly and yields gauge-parameter-independent results for
physical $S$-matrix elements up absorptive parts related to the
presence of external unstable particles.  Moreover, it is
symmetric with respect to the fermion generations and, at least
for the  non-absorptive parts, avoids unphysical singularities in
the limit of degenerate quark masses.

\appendix

\section{Renormalization of the quark-mixing matrix in the presence of
LSZ factors including absorptive parts}

In \citere{Espriu:2002xv} it has been argued that in the case of
unstable external fermions different sets of LSZ factors
have to be introduced for incoming and
outgoing fermions in order to obtain gauge-parameter-independent
amplitudes. Although this approach has a certain appeal, it is still far
from a description of amplitudes for unstable particles.
The purpose of this appendix is to investigate the renormalization of the
QMM in this approach.

The matrix-valued LSZ factors for all left- and
right-handed incoming fermions and outgoing anti-fermions are denoted
by $\delta Z^\qL$ and $\delta Z^\qR$ and those for the outgoing
fermions and incoming anti-fermions by $\delta \Zbar^\qL$ and
$\delta \Zbar^\qR$. Similarly, the LSZ factor for the incoming $\PW^+$
boson and outgoing $\PW^-$ boson is given by $\delta Z_\PW$ and the
one for the incoming $\PW^-$ boson and outgoing $\PW^+$ boson by
$\delta \Zbar_\PW$. These LSZ factors are suitable for $S$-matrix
elements and involve enough freedom to include all absorptive parts.
However, they should not be used as field renormalization constants in the
renormalized Lagrangian, since they would violate its hermiticity. In
the following, the constants $\delta Z$ and $\delta \Zbar$ can be
understood to include both the LSZ factors and field-renormalization
constants or only the LSZ factors assuming that field renormalization
has already been performed.

The LSZ factors are fixed by imposing on-shell conditions for incoming
and outgoing fermions \cite{Espriu:2002xv}.  From the resulting
explicit expressions for $\delta Z^\qLR$ and $\delta \Zbar^\qLR$
in terms of self-energies given in \citere{Espriu:2002xv} and those of
the self-energies we find the relations
\begin{equation}
\label{a:Zsymmetry}
\delta \Zbar^\qLR = \left.(\delta Z^\qLR)^\rT\right|_{V\to V^*},
\end{equation}
or equivalently
\begin{equation}
\label{a:Zhermiticity}
\Retilde \delta \Zbar^\qLR = \Retilde (\delta Z^\qLR)^\dagger,
\qquad
\Imtilde \delta \Zbar^\qLR = -\Imtilde (\delta Z^\qLR)^\dagger
\end{equation}
using ${\Retilde}$ and ${\Imtilde}$ as defined in
\citere{Denner:1993kt}, which acts only on the loop integrals but not
on the QMM, \ie, ${\Retilde}$ projects on dispersive parts and
${\Imtilde}$ on absorptive parts.  As a consequence of
\refeq{a:Zhermiticity}, the barred matrices $\delta \Zbar^\qL$ and
$\delta \Zbar^\qR$ agree with $(\delta Z^\qL)^\dagger$ and $(\delta
Z^\qR)^\dagger$ in the dispersive parts but have a different sign for
the absorptive parts. Analogous relations hold between $\delta
Z_\PW^*$ and $\delta \Zbar_\PW$.

In the approach of \citere{Espriu:2002xv} the quantum corrections to
the decays $\PW^+ \to \Pu_i \bar\Pd_j$ and
$\bar\Pu_i\to\PW^-\bar\Pd_j$ are given by \refeqs{eq:Mcorrections},
\refeqf{eq:formfactor}, and \refeqf{eq:Wudvert} with $(\delta
Z^{\uL})^\dagger$ replaced by $\delta \Zbar^{\uL}$. Using the physical
renormalization condition \refeq{eq:rencond3} leads to
\begin{eqnarray}
\label{a:CPresultconventional}
\delta \Vah &=&
- \frac{1}{2} \left[\delta \Zbar^\uL - (\delta \Zbar^\uL)^\dagger
  \right] V
- \frac{1}{2} V \left[\delta Z^\dL - (\delta Z^\dL)^\dagger
  \right]
\nonumber \\ && {}
- \frac{1}{2} \left(\delta Z_\PW - \delta Z_\PW^*\right) V
- \frac{1}{2} \left[\delta F_{\loop,1}^{-} - V (\delta F_{\loop,1}^-)^\dagger V \right]
\end{eqnarray}
instead of \refeq{eq:CPresultconventional}.
This renormalization condition is equivalent to the one given by Zhou
in \citere{Zhou:2003gb}.

Using the decays $\PW^- \to \bar\Pu_i\Pd_j$ 
and $\Pu_i \to\PW^+ \Pd_j$
with an analogous renormalization condition, yields instead
\begin{eqnarray}
\label{a:CPresultconventional2}
\delta \Vah  & = &
- \frac{1}{2} \left[(\delta Z^\uL)^\dagger - \delta Z^\uL\right] V
- \frac{1}{2} V \left[(\delta \Zbar^\dL)^\dagger - \delta \Zbar^\dL
  \right]
\nonumber \\ && {}
- \frac{1}{2} \left(\delta \Zbar_\PW^*  - \delta \Zbar_\PW \right) V
- \frac{1}{2} \left[(\delta G_{\loop,1}^-)^\dagger - V \delta G_{\loop,1}^- V \right],
\end{eqnarray}
where $\delta G_{\loop,1}^-$ is defined in analogy to $\delta
F_{\loop,1}^-$.
These matrices are related by
\begin{equation}
\label{a:FGrel}
\delta G_{\loop,1}^- = \left.(\delta F_{\loop,1}^-)^\rT \right|_{V\to V^*},
\end{equation}
or equivalently
\begin{equation}
\label{a:FGrel2}
\Retilde\delta G^-_{\loop,1} = \Retilde(\delta F^-_{\loop,1})^\dagger,\qquad
\Imtilde\delta G^-_{\loop,1} = -\Imtilde(\delta F^-_{\loop,1})^\dagger,
\end{equation}
as can be seen from the structure of the explicit expressions. The
counterterms \refeq{a:CPresultconventional} and
\refeq{a:CPresultconventional2} involve the same dispersive parts
but opposite absorptive parts.  In the case of CP conservation the
renormalization conditions \refeq{a:CPresultconventional} and
\refeq{a:CPresultconventional2} violate the orthogonality and
reality of the QMM.  These drawbacks can be cured by omitting all
absorptive parts in the counterterms  for $\delta \Vah$. In this
case the expressions \refeq{a:CPresultconventional} and
\refeq{a:CPresultconventional2} become equivalent, and the
counterterm simplifies to
\begin{eqnarray}
\label{a:CPresultconventional3}
\delta \Vah &=& -\Retilde \Bigg\{
  \frac{1}{2} \left[(\delta Z^\uL)^\dagger - \delta Z^\uL
  \right] V
+ \frac{1}{2} V  \left[\delta Z^\dL - (\delta  Z^{\rd,\rL})^\dagger
  \right]
\nonumber \\ && {}
+ \frac{1}{2} \left[\delta F_{\loop,1}^- - V (\delta F_{\loop,1}^-)^\dagger
  V \right]\Bigg\}.
\end{eqnarray}
Alternatively, this result can be directly obtained from the renormalization
condition
\begin{equation}
[F_1^-+(G_1^-)^\dagger]V^\dagger = V[(F_1^-)^\dagger+G_1^-],
\end{equation}
where $G_{\loop,1}^-$ is the matrix replacing $F_{\loop,1}^-$ in these
decays, and the relations
\refeq{a:Zhermiticity} and \refeq{a:FGrel2} without the need to use
the $\Retilde$ prescription.

As a consequence, also in this approach the most natural strategy
is to discard all absorptive parts in the renormalization
constants of the QMM. Then, the counterterms for $\delta \Vah$ are
the same as those 
introduced in \refeq{eq:CPresultconventional} with the same merits and
drawbacks.

Finally, we note that one could construct a counterterm that yields
$S$-matrix elements for the decays $\PW^+ \to \Pu_i \bar\Pd_j$ and
$\bar\Pt\to\PW^-\bar\Pd_j$ where the limit of degenerate fermion masses is
approached smoothly by including appropriate absorptive parts of the
LSZ factors in \refeq{a:CPresultconventional3} as
\begin{eqnarray}
\label{a:CPresultnonsing}
\delta \Vah &=& -
  \frac{1}{2} \left[\delta \Zbar^\uL - \delta Z^\uL
  \right] V
- \frac{1}{2} V  \left[\delta Z^\dL - \delta  \Zbar^{\rd,\rL}
  \right]
- \frac{1}{2} \left[\delta F_{\loop,1}^- - V (\delta F_{\loop,1}^-)^\dagger
  V \right].\nln
\end{eqnarray}
With this counterterm only the combinations $\delta \Zbar^\uL + \delta
Z^\uL$ and $\delta \Zbar^\dL + \delta Z^\dL$ that are non-singular for
degenerate fermion masses appear in the renormalized $S$-matrix
elements for $\PW^+ \to \Pu_i \bar\Pd_j$ and
$\bar\Pu_i\to\PW^-\bar\Pd_j$.  However, the absorptive parts in
\refeq{a:CPresultnonsing} violate the unitarity of the renormalized
QMM and are thus not admissible. Moreover, even for these counterterms
the $S$-matrix elements for the decays $\PW^- \to \bar\Pu_i\Pd_j$ 
and $\Pu_i\to\PW^+\Pd_j$ are not smooth in the degenerate-quark-mass
limit.

\section*{Acknowledgement}

This work was supported in part by the Swiss Bundesamt f\"ur Bildung und
Wissenschaft and by the European Union under contract
HPRN-CT-2000-00149.

\end{document}